\documentstyle[12pt]{article}
\begin{document}
\title{Locally Anisotropic Interactions:\\
I. Nonlinear Connections in Higher Order Anisotropic Superspaces}
\author{Sergiu I.\ Vacaru }
\maketitle

\centerline{\noindent{\em Institute of Applied Physics, Academy of Sciences,}}
\centerline{\noindent{\em 5 Academy str., Chi\c sin\v au 2028,
Republic of Moldova}} \vskip10pt
\centerline{\noindent{ Fax: 011-3732-738149, E-mail: lises@cc.acad.md}}
\begin{abstract}
Higher order anisotropic superspaces are constructed as generalized vector
superbundles provided with compatible nonlinear connection, distinguished
connection and metric structures.
\end{abstract}

\section{Introduction}

The differential supergeometry have been formulated with the aim of getting
a geometric framework for the supersymmetric field theories (see the theory
of graded manifolds \cite{ber,lei,leim,kon}, the theory of supermanifolds
\cite{dew,rog80,bar,jad} and, for detailed considerations of geometric and
topological aspects of supermanifolds and formulation of superanalysis, \cite
{cia,bru,man,hoy,vla,vol}). In this paper we apply the supergeometric
formalism for a study of a new class of (higher order anisotropic)
superspaces.

The concept of local anisotropy is largely used in some divisions of
theoretical and mathematical physics \cite{vlas,in,ish,mk} (see also
possible applications in physics and biology in \cite{az94,am}). The first
models of locally anisotropic (la) spaces (la--spaces) have been proposed by
P.Finsler \cite{fin} and E.Cartan \cite{car35} (early approaches and modern
treatments of Finsler geometry and its extensions can be found, for
instance, in \cite{run,asa,asa88,mat}). In our works \cite
{vjmp,vlasg,vsp96,vg,voa} we try to formulate the geometry
of la-spaces in a manner as to include both variants of Finsler and
Lagrange, in general supersymmetric, extensions and higher dimensional
Kaluza--Klein (super)spaces as well to propose general principles and
methods of construction of models of classical and quantum field
interactions and stochastic processes on spaces with generic anisotropy.

We cite here the works \cite{bej90,bej91} by A. Bejancu where a new
viewpoint on differential geometry of supermanifolds is considered. The
author introduced the nonlinear connection (N--connection) structure and
developed a corresponding distinguished by N--connection supertensor
covariant differential calculus in the frame of De Witt \cite{dew} approach
to supermanifolds in the framework of the geometry of superbundles with
typical fibres parametrized by noncommutative coordinates. This was the
first example of superspace with local anisotropy. In our turn we have given
a general definition of locally anisotropic superspaces (la--superspaces)
\cite{vlasg}. We note that in some of our supersymmetric generalizations we
are inspired by the R. Miron, M. Anastasiei and Gh. Atanasiu works on the
geometry of nonlinear connections in vector bundles and higher order
Lagrange spaces \cite{ma87,ma94,mirata}.

In this work we shall formulate the theory of higher order vector
superbundles provided with nonlinear and distinguished connections and
metric structures (a generalized model of la--superspaces). Such
superbundles contain as particular cases the supersymmetric extensions and
various higher order prolongations of Riemann, Finsler and Lagrange spaces.

The paper is organized as follows: Section 2 is a brief review on
supermanifolds and superbundles. An introduction into the geometry of higher
order distinguished vector superbundles is presented in Section 3. Section 4
and 5 deals respectively with the geometry of nonlinear and linear
distinguished connections in vector superbundles and in distinguished vector
superbundles. Concluding remarks and discussion are contained in Section 6.

\section{Supermanifolds and Superbundles}

In this section we establish the necessary terminology on supermanifolds
(s--manifolds) \cite{dew,rog80,rog81,jad,vla,hoy,man,bar,bru,cia,hoy}. Here
we note that a number of different approaches to supermanifolds are broadly
equivalent for local considerations. For simplicity, we shall restrict our
study only with geometric constructions on locally trivial superspaces.

To build up s--manifolds (see \cite{rog80,jad,vla}) one uses as basic
structures \ Grassmann algebra and Banach space. A Grassmann algebra is
introduced as a real associative algebra $\Lambda $ (with unity) possessing
a finite (canonical) set of anticommutative generators $\beta _{\hat A}$, ${{%
[{\beta _{\hat A}},{\beta _{\hat B}}]}_{+}}={{\beta _{\hat A}}}{{\beta
_{\hat C}}}+{{\beta _{\hat C}}}{{\beta _{\hat A}}}=0$, where ${{\hat A},{%
\hat B},...}=1,2,...,{\hat L}$. In this case it is also defined a ${Z_2}$%
-graded commutative algebra ${{\Lambda }_0}+{{\Lambda }_1}$, whose even part
${{\Lambda }_0}$ (odd part ${{\Lambda }_1}$) is a ${2^{{\hat L}-1}}$%
--dimensional real vector space of even (odd) products of generators ${\beta
}_{\hat A}$.After setting ${{\Lambda }_0}={\cal R}+{{\Lambda }_0}^{\prime }$%
, where ${\cal R}$ is the real number field and ${{\Lambda }_0}^{\prime }$
is the subspace of ${\Lambda }$ consisting of nilpotent elements, the
projections ${\sigma }:{\Lambda }\to {\cal R}$ and $s:{\Lambda }\to {{%
\Lambda }_0}^{\prime }$ are called, respectively, the body and soul maps.

A Grassmann algebra can be provided with both structures of a Banach algebra
and Euclidean topological space by the norm \cite{rog80}
$$
{\Vert }{\xi }{\Vert }={{\Sigma }_{{\hat A}_i}}{|}a^{{{\hat A}_1}...{{\hat A}%
_k}}{|},{\xi }={{\Sigma }_{r=0}^{\hat L}}a^{{{\hat A}_1}...{{\hat A}_r}}{{%
\beta }_{{\hat A}_1}}...{{\beta }_{{\hat A}_r}}.
$$
A superspace is introduced as a product
$$
{\Lambda }^{n,k}={\underbrace{{{\Lambda }_0}{\times }...{\times }{{\Lambda }%
_0}}_n{\times }{\underbrace{{{\Lambda }_1}{\times }...{\times }{{\Lambda }_1}%
}_k}}
$$
which is the $\Lambda $-envelope of a $Z_2$-graded vector space ${V^{n,k}}={%
V_0}{\otimes }{V_1}={{\cal R}^n}\oplus {{\cal R}^k}$ is obtained by
multiplication of even (odd) vectors of $V$ on even (odd) elements of ${%
\Lambda }$. The superspace (as the ${\Lambda }$-envelope) posses $(n+k)$
basis vectors $\{{\hat {{\beta }_i}},{\quad }i=0,1,...,n-1,$ and ${\quad }{{%
\beta }_{\hat i}},{\quad }{\hat i}=1,2,...k\}$. Coordinates of even (odd)
elements of $V^{n,k}$ are even (odd) elements of $\Lambda $. We can consider
equivalently a superspace $V^{n,k}$ as a $({2^{{\hat L}-1}})(n+k)$%
-dimensional real vector spaces with a basis $\{{{\hat \beta }_{i({\Lambda }%
)}},{{\beta }_{{\hat i}({\Lambda })}}\}$.

Functions of superspaces, differentiation with respect to Grassmann
coordinates, supersmooth (superanalytic) functions and mappings are
introduced by analogy with the ordinary case, but with a glance to certain
specificity caused by changing of real (or complex) number field into
Grassmann algebra $\Lambda $. Here we remark that functions on a superspace $%
{\Lambda }^{n,k}$ which takes values in Grassmann algebra can be considered
as mappings of the space ${\cal R}^{{({2^{({\hat L}-1)}})}{(n+k)}}$ into the
space ${\cal R}^{2{\hat L}}$. Functions differentiable on Grassmann
coordinates can be rewritten via derivatives on real coordinates, which obey
a generalized form of Cauchy-Riemann conditions.

A $(n,k)$-dimensional s-manifold $\tilde M$ can be defined as a Banach
manifold (see, for example, \cite{len}) modeled on ${\Lambda }^{n,k}$
endowed with an atlas ${\psi }={\{}{U_{(i)}},{{\psi }_{(i)}}:{U_{(i)}}\to {{%
\Lambda }^{n,k}},(i)\in J{\}}$ whose transition functions ${\psi }_{(i)}$
are supersmooth \cite{rog80,jad}. Instead of supersmooth functions we can
use $G^\infty $-functions \cite{rog80} and introduce $G^\infty $%
-supermanifolds ($G^\infty $ denotes the class of superdifferentiable
functions). The local structure of a $G^\infty $-supermanifold is built very
much as on a $C^\infty $-manifold. Just as a vector field on a $n$%
-dimensional $C^\infty $-manifold written locally as%
$$
{\Sigma }_{i=0}^{n-1}{\quad }{f_i}{(x^j)}{\frac{{\partial }}{{{\partial }x^i}%
}},
$$
where $f_i$ are $C^\infty $-functions, a vector field on an $(n,k)$%
-dimensional $G^\infty $--su\-per\-ma\-ni\-fold $\tilde M$ can be expressed
locally on an open region $U{\subset }\tilde M$ as
$$
{\Sigma }_{I=0}^{n-1+k}{\quad }{f_I}{(x^J)}{\frac{{\partial }}{{{\partial }%
x^I}}}= {\Sigma }_{i=0}^{n-1}{\quad }{f_i}{(x^j,{{\theta }^{\hat j}})}{\frac{%
{\partial }}{{{\partial }x^i}}}+{\Sigma }_{{\hat i}=1}^k{\quad }{f_{\hat i}}{%
(x^j,{{\theta }^{\hat j}})}{\frac \partial {\partial {{\theta }^{\hat i}}}},
$$
where $x=({\hat x},{\theta })=\{{x^I}=({{\hat x}^i},{\theta }^{\hat i})\}$
are local (even, odd) coordinates. We shall use indices $I=(i,{\hat i}),J=(j,%
{\hat j}),K=(k,{\hat k}),...$ for geometric objects on $\tilde M$. A vector
field on $U$ is an element $X{\subset }End[{G^\infty }(U)]$ (we can also
consider supersmooth functions instead of $G^\infty $-functions) such that
$$
X(fg)=(Xf)g+{(-)}^{{\mid }f{\mid }{\mid }X{\mid }}fXg,
$$
for all $f,g$ in $G^\infty (U)$, and
$$
X(af)={(-)}^{{\mid }X{\mid }{\mid }a{\mid }}aXf,
$$
where ${\mid }X{\mid }$ and ${\mid }a{\mid }$ denote correspondingly the
parity $(=0,1)$ of values $X$ and $a$ and for simplicity in this work we
shall write ${(-)}^{{\mid }f{\mid }{\mid }X{\mid }}$ instead of ${(-1)}^{{%
\mid }f{\mid }{\mid }X{\mid }}.$

A super Lie group (sl-group) \cite{rog81} is both an abstract group and a
s-manifold, provided that the group composition law fulfills a suitable
smoothness condition (i.e. to be superanalytic, for short, $sa$ \cite{jad}).

In our further considerations we shall use the group of automorphisms of ${%
\Lambda}^{(n,k)}$, denoted as $GL(n,k,{\Lambda})$, which can be parametrized
as the super Lie group of invertible matrices
$$
Q={\left(
\begin{array}{cc}
A & B \\
C & D
\end{array}
\right) } ,%
$$
where A and D are respectively $(n{\times}n)$ and $(k{\times}k)$ matrices
consisting of even Grassmann elements and B and C are rectangular matrices
consisting of odd Grassmann elements. A matrix Q is invertible as soon as
maps ${\sigma}A$ and ${\sigma}D$ are invertible matrices. A sl-group
represents an ordinary Lie group included in the group of linear transforms $%
GL(2^{{\hat L}-1}(n+k),{\cal R})$. For matrices of type Q one defines \cite
{ber,lei,leim} the superdeterminant, $sdetQ$, supertrace, $strQ$, and
superrank, $srankQ$.

A Lie superalgebra (sl-algebra) is a $Z_2$-graded algebra $A={A_0}\oplus A_1$
endowed with product $[,\}$ satisfying the following properties:
$$
[I,I^{\prime }\}=-{(-)}^{{\mid }I{\mid }{\mid }I^{\prime }{\mid }}[I^{\prime
},I\},
$$
$$
[I,[I^{\prime },I^{\prime \prime }\}\}=[[I,I^{\prime }\},I^{\prime \prime
}\}+{(-)}^{{\mid }I{\mid }{\mid }I^{\prime }{\mid }}[I^{\prime }[I,I^{\prime
\prime }\}\},
$$
$I{\in }A_{{\mid }I{\mid }},{\quad }I^{\prime }{\in }A_{{\mid }I^{\prime }{%
\mid }}$, where ${\mid }I{\mid },{\mid }I^{\prime }{\mid }=0,1$ enumerates,
respectively, the possible parity of elements $I,I^{\prime }$. The even part
$A_0$ of a sl-algebra is a usual Lie algebra and the odd part $A_1$ is a
representation of this Lie algebra. This enables us to classify sl--algebras
following the Lie algebra classification \cite{kac}. We also point out that
irreducible linear representations of Lie superalgebra A are realized in $%
Z_2 $-graded vector spaces by matrices $\left(
\begin{array}{cc}
A & 0 \\
0 & D
\end{array}
\right) $ for even elements and $\left(
\begin{array}{cc}
0 & B \\
C & 0
\end{array}
\right) $ for odd elements and that, roughly speaking, A is a superalgebra
of generators of a sl-group.

A sl--module $W$ (graded Lie module) \cite{rog80} is introduced as a $Z_2$%
-graded left $\Lambda $-module endowed with a product $[,\}$ which satisfies
the graded Jacobi identity and makes $W$ into a graded-anticommuta\-ti\-ve
Banach algebra over $\Lambda $. One calls the Lie module {\cal G} the set of
the left-invariant derivatives of a sl-group $G$.

The tangent superbundle (ts-bundle) $T\tilde M$ over a s-manifold $\tilde M$%
, ${\pi }:T\tilde M\to {\tilde M}$ is constructed in a usual manner (see,
for instance, \cite{len}) by taking as the typical fibre the superspace ${%
\Lambda }^{n,k}$ and as the structure group the group of automorphisms, i.e.
the sl-group $GL(n,k,{\Lambda }).$

Let us denote by ${\cal F}$ a vector superspace (vs-space) of dimension $%
(m,l)$ (with respect to a chosen base we parametrize an element $y\in {\cal E%
}$ as $y=({\hat y},\zeta )=\{{y^A}=({\hat {y^a}},{\zeta }^{\hat a})\}$,
where $a=1,2,...,m$ and ${\hat a}=1,2,...,l$). We shall use indices $A=(a,{%
\hat a}),B=(b,{\hat b}),...$ for objects on vs-spaces. A vector superbundle
(vs-bundle) $\tilde {{\cal E}}$ over base $\tilde M$ with total superspace $%
\tilde E$, standard fibre ${\hat {{\cal F}}}$ and surjective projection ${{%
\pi }_E}:\tilde E{\to }\tilde M$ is defined (see details and variants in
\cite{bru,vol}) as in the case of ordinary manifolds (see, for instance,
\cite{len,ma87,ma94}). A section of $\tilde {{\cal E}}$ is a supersmooth map
$s:U{\to }\tilde E$ such that ${{\pi }_E}{\cdot }s=id_U.$

A subbundle of ${\tilde {{\cal E}}}$ is a triple $(\tilde {{\cal B}%
},f,f^{\prime})$, where $\tilde {{\cal B}}$ is a vs-bundle on $\tilde M$,
maps $f: \tilde {{\cal B}} {\to} \tilde {{\cal E}}$ and $f^{\prime}: \tilde M%
{\to} \tilde M$ are supersmooth, and $(i) {\quad}{{\pi}_E}{\circ}f=f^{\prime}%
{\circ}{{\pi}_B};$ $(ii) {\quad} f:{\pi}^{-1}_B {(x)} {\to} {\pi}^{-1}_E {%
\circ} f^{\prime}(x)$ is a vs-space homomorphism.

We denote by
$$
u=(x,y)=({\hat x},{\theta },{\hat y},{\zeta })=\{u^\alpha =(x^I,y^A)=({{\hat
x}^i},{\theta }^{\hat i},{{\hat y}^a},{\zeta }^{\hat a})=({{\hat x}^i}%
,x^{\hat i},{{\hat y}^a},y^{\hat a})\}
$$
the local coordinates in ${\tilde {{\cal E}}}$ and write their
transformations as
$$
x^{I^{\prime }}=x^{I^{\prime }}({x^I}),{\quad }srank({\frac{{\partial }%
x^{I^{\prime }}}{{\partial }x^I}})=(n,k),\eqno(1)
$$
$y^{A^{\prime }}=Y_A^{A^{\prime }}(x){y}^A,$ where $Y_A^{A^{\prime }}(x){\in
}G(m,l,\Lambda ).$

For local coordinates and geometric objects on ts-bundle $T\tilde M$ we
shall not distinguish indices of coordinates on the base and in the fibre
and write, for instance,
$$
u=(x,y)=({\hat x},{\theta },{\hat y},{\zeta })=\{u^\alpha =({x^I},{y^I})=({{%
\hat x}^i},{\theta }^{\hat i},{{\hat y}^i},{\zeta }^{\hat i})=({{\hat x}^i}%
,x^{\hat i},{{\hat y}^i},y^{\hat i})\}.
$$
We shall use general Greek indices on both type of vs- and ts--bundles.

\section{Distinguished Vector Superbundles}

Some recent considerations in mathematical physics are based on the
so--called k--jet spaces (see, for instance, \cite{sau,sar,asa89}). In order
to formulate a systematic theory of connections and of geometric structures
on k--jet bundles, in a manner following the approaches \cite{yan} and \cite
{ma87,ma94}, R. Miron and Gh. Atanasiu \cite{mirata} introduced the concept
of k--osculator bundle for which a fiber of k-jets is changed into a
k--osculator fiber representing an element of k--order curve. Such
considerations are connected with geometric constructions on tangent bundles
of higher order. On the other hand for developments in modern supersymmetric
Kaluza--Klein theories (see, for instance, \cite{sal}) a substantial
interest would present a variant of ''osculator'' space for which the higher
order tangent s-space distributions are of different dimensions. This
section is devoted to the definition of such type distinguished vector
superbundle spaces.

A vector superspace ${\cal F}^{<z>}$ of dimension $(m,l)$ is a distinguished
vector superspace (dvs--space) if it is decomposed into an invariant
oriented direct sum ${\cal F}^{<z>}={\cal F}_{(1)}\oplus {\cal F}%
_{(2)}\oplus ...\oplus {\cal F}_{(z)}$ of vs--spaces ${\cal F}_{(p)},\dim
{\cal F}_{(p)}=(m_{(p)},l_{(p)}),$ where $(p)=(1),(2),...,(z),%
\sum_{p=1}^{p=z}m_{(p)}=m,\sum_{p=1}^{p=z}l_{(p)}=l.$

Coordinates on ${\cal F}^{<p>}$ will be parametrized as
$$
{y}^{<p>}=(y_{(1)},y_{(2)},...,y_{(p)})=({\hat y}_{(1)},{\zeta }_{(1)},{\hat
y}_{(2)},{\zeta }_{(2)},...,{\hat y}_{(p)},{\zeta }_{(p)})=
$$
$$
\{y^{<A>}=({{\hat y}^{<a>}},{\zeta }^{<\hat a>})=({{\hat y}^{<a>}},y^{<\hat
a>})\},
$$
where bracketed indices are correspondingly split on ${\cal F}_{(p)}$%
--components:
$$
<A>=\left( A_{(1)},A_{(2)},...,A_{(p)}\right)
,<a>=(a_{(1)},a_{(2)},...,a_{(p)})
$$
$$
\mbox{ and }<\widehat{a}>=(\widehat{a}_{(1)}\widehat{a}_{(2)},...,\widehat{a}%
_{(p)}),\eqno(2)
$$
For simplicity, we shall also write (2) as $<A>=\left(
A_1,A_2,...,A_p\right) ,<a>=(a_1,a_2,...,a_p)$ and $<\widehat{a}>=(\widehat{a%
}_1\widehat{a}_2,...,\widehat{a}_p)$ if this will give not rise to
ambiguities.

A distinguished vector superbundle (dvs--bundle)\\ $\widetilde{{\cal E}}%
^{<z>}=(\tilde E^{<z>},\pi ^{<d>},{\cal F}^{<d>},\tilde M),$ with surjective
projection $\pi ^{<z>}:\tilde E^{<z>}\rightarrow \tilde M,$ where $\tilde M$
and $\tilde E^{<z>}$ are respectively base and total s--spaces and the
dv--space ${\cal F}^{<z>}$ is the standard fibre, is defined in a usual
manner (see correspondingly \cite{bru,cia,bar,vol} on vector superbundles
and \cite{len,ma94,ma87} on vector bundles).

A dvs--bundle $\widetilde{{\cal E}}^{<z>}$ is constructed as an oriented set
of vs--bundles $\pi ^{<p>}:\tilde E^{<p>}\rightarrow \tilde E^{<p-1>}$ (with
typical fibers ${\cal F}^{<p>},p=1,2,...,z);$ $\tilde E^{<0>}=\tilde M.$ We
shall use index $z~(p)$ as to denote the total (intermediate) numbers of
consequent vs--bundle coverings of $\tilde M.$

Local coordinates on $\widetilde{{\cal E}}^{<p>}$ are denoted as%
$$
u_{(p)}=(x,y_{<p>})=(x,y_{(1)},y_{(2)},...,y_{(p)})=
$$
$$
({\hat x},{\theta },{\hat y}_{<p>},{\zeta }_{<p>})=({\hat x},{\theta },{\hat
y}_{(1)},{\zeta }_{(1)},{\hat y}_{(2)},{\zeta }_{(2)},...,{\hat y}_{(p)},{%
\zeta }_{(p)})=
$$
$$
\{u^{<\alpha >}=(x^I,y^{<A>})=({{\hat x}^i},{\theta }^{\hat i},{{\hat y}%
^{<a>}},{\zeta }^{<\hat a>})=({{\hat x}^i},x^{\hat i},{{\hat y}^{<a>}}%
,y^{<\hat a>})\}=...
$$
(in our further considerations we shall consider different variants of
splitting indices of geometric objects).

Instead of (1) the coordinate transforms for dvs--bundles\\ $\{u^{<\alpha
>}=(x^I,y^{<A>})\}\rightarrow \{u^{<\alpha ^{\prime }>}=(x^{I^{\prime
}},y^{<A^{\prime }>})\}$ are given by recurrent maps:%
$$
x^{I^{\prime }}=x^{I^{\prime }}({x^I}),{\quad }srank({\frac{{\partial }%
x^{I^{\prime }}}{{\partial }x^I}})=(n,k),\eqno(3)
$$
$$
y_{(1)}^{A_1^{\prime }}=K_{A_1}^{A_1^{\prime }}(x){y}%
_{(1)}^{A_1},K_{A_1}^{A_1^{\prime }}(x){\in }G(m_{(1)},l_{(1)},\Lambda ),
$$
$$
..................................................
$$
$$
y_{(p)}^{A_p^{\prime }}=K_{A_p}^{A_p^{\prime }}(u_{(p-1)}){y}%
_{(p)}^{A_p},K_{A_p}^{A_p^{\prime }}(u_{(p-1)}){\in }G(m_{(p)},l_{(p)},%
\Lambda ),
$$
$$
.................................................
$$
$$
y_{(z)}^{A_z^{\prime }}=K_{A_z}^{A_z^{\prime }}(u_{(z-1)}){y}%
_{(z)}^{A_z},K_{A_z}^{A_z^{\prime }}(u_{(z-1)}){\in }G(m_{(z)},l_{(z)},%
\Lambda ).
$$
In brief we write transforms (3) as
$$
x^{I^{\prime }}=x^{I^{\prime }}(x^I),~y^{<A^{\prime }>}=K_{<A>}^{<A^{\prime
}>}y^{<A>}.
$$
More generally, we shall consider matrices $K_{<\alpha >}^{<\alpha ^{\prime
}>}=(K_I^{I^{\prime }},K_{<A>}^{<A^{\prime }>}),$ where $K_I^{I^{\prime
}}\doteq \frac{\partial x^{I^{\prime }}}{\partial x^I}.$

In consequence the local coordinate bases of the module of ds-vector fields
 on $\widetilde{{\cal E}}^{<z>},\ $\  $%
\Theta (\widetilde{{\cal E}}^{<z>}),$
$$
\partial _{<\alpha >}=(\partial _I,\partial _{<A>})=(\partial _I,\partial
_{(A_1)},\partial _{(A_2)},...,\partial _{(A_z)})=
$$
$$
\frac \partial {\partial u^{<\alpha >}}=(\frac \partial {\partial x^I},\frac
\partial {\partial y_{(1)}^{A_1}},\frac \partial {\partial
y_{(2)}^{A_2}},...,\frac \partial {\partial y_{(z)}^{A_z}})\eqno(4)
$$
(the dual coordinate bases are denoted as%
$$
d^{<\alpha >}=(d^I,d^{<A>})=(d^I,d^{(A_1)},d^{(A_2)},...,d^{(A_z)})=
$$
$$
du^{<\alpha >}=(dx^I,dy^{(A_1)},dy^{(A_2)},...,dy^{(A_z)})\quad )\eqno(5)
$$
are transformed as%
$$
\partial _{<\alpha >}=(\partial _I,\partial _{<A>})=(\partial _I,\partial
_{(A_1)},\partial _{(A_2)},...,\partial _{(A_z)})\rightarrow \partial
_{<\alpha >}=
$$
$$
(\partial _I,\partial _{<A>})=(\partial _I,\partial _{(A_1)},\partial
_{(A_2)},...,\partial _{(A_z)})
$$
$$
\frac \partial {\partial x^I}=K_I^{I^{\prime }}\frac \partial {\partial
x^{I^{\prime }}}+Y_{(1,0)I}^{A_1^{\prime }}\frac \partial {\partial
y_{(1)}^{A_1^{\prime }}}+Y_{(2,0)I}^{A_2^{\prime }}\frac \partial {\partial
y_{(2)}^{A_2^{\prime }}}+...+Y_{(z,0)I}^{A_z^{\prime }}\frac \partial
{\partial y_{(z)}^{A_z^{\prime }}},\eqno(6)
$$
$$
\frac \partial {\partial y_{(1)}^{A_1}}=K_{A_1}^{A_1^{\prime }}\frac
\partial {\partial y_{(1)}^{A_1^{\prime }}}+Y_{(2,1)A_1}^{A_2^{\prime
}}\frac \partial {\partial y_{(2)}^{A_2^{\prime
}}}+...+Y_{(z,1)A_1}^{A_z^{\prime }}\frac \partial {\partial
y_{(z)}^{A_z^{\prime }}},
$$
$$
\frac \partial {\partial y_{(2)}^{A_2}}=K_{A_2}^{A_2^{\prime }}\frac
\partial {\partial y_{(2)}^{A_2^{\prime }}}+Y_{(3,2)A_2}^{A_3^{\prime
}}\frac \partial {\partial y_{(3)}^{A_3^{\prime
}}}+...+Y_{(z,2)A_2}^{A_z^{\prime }}\frac \partial {\partial
y_{(z)}^{A_z^{\prime }}},
$$
$$
........................................................
$$
$$
\frac \partial {\partial y_{(z-1)}^{A_{z-1}}}=K_{A_{z-1}}^{A_{z-1}^{\prime
}}\frac \partial {\partial y_{(z-1)}^{A_{z-1}^{\prime
}}}+Y_{(z,z-1)A_{s-1}}^{A_z^{\prime }}\frac \partial {\partial
y_{(z)}^{A_z^{\prime }}},
$$
$$
\frac \partial {\partial y_{(z)}^{A_z}}=K_{A_z}^{A_z^{\prime }}\frac
\partial {\partial y_{(z)}^{A_z^{\prime }}}.
$$

$Y$--matrices from (6) are partial derivations of corresponding combinations
of $K$--coefficients from coordinate transforms (3),
$$
Y_{A_f}^{A_p^{\prime }}=\frac{\partial (K_{A_p}^{A_p^{\prime }}~y^{A_p})}{%
\partial y^{A_f}},~f<p.
$$

In brief we denote respectively ds-coordinate transforms of coordinate bases
(4)\ and (5)\ as%
$$
\partial _{<\alpha >}=(K_{<\alpha >}^{<\alpha ^{\prime }>}+Y_{<\alpha
>}^{<\alpha ^{\prime }>})~\partial _{<\alpha ^{\prime }>}\mbox{ and }%
~d^{<\alpha >}=(K_{<\alpha ^{\prime }>}^{<\alpha >}+Y_{<\alpha ^{\prime
}>}^{<\alpha >})d^{<\alpha ^{\prime }>},
$$
where matrix $K_{<\alpha >}^{<\alpha ^{\prime }>}$, its s-inverse $%
K_{<\alpha ^{\prime }>}^{<\alpha >}$, as well $Y_{<\alpha >}^{<\alpha
^{\prime }>}$ and $Y_{<\alpha ^{\prime }>}^{<\alpha >}$ are paramet\-riz\-ed
according to (6). In order to illustrate geometric properties of some of our
transforms it is useful to introduce matrix operators and to consider in
explicit form the parametrizations of matrices under consideration. For
instance, in operator form the transforms (6)
$$
{\bf \partial =}\widehat{{\bf Y}}{\bf \partial }^{\prime },
$$
are characterized by matrices of type
$$
{\bf \partial =}\partial _{<\alpha >}=\left(
\begin{array}{c}
\partial _I \\
\partial _{A_1} \\
\partial _{A_2} \\
... \\
\partial _{A_z}
\end{array}
\right) =\left(
\begin{array}{c}
\frac \partial {\partial x^I} \\
\frac \partial {\partial y_{(1)}^{A_1}} \\
\frac \partial {\partial y_{(2)}^{A_2}} \\
... \\
\frac \partial {\partial y_{(z)}^{A_z}}
\end{array}
\right) ,{\bf \partial }^{\prime }{\bf =}\partial _{<\alpha ^{\prime
}>}=\left(
\begin{array}{c}
\partial _{I^{\prime }} \\
\partial _{A_1^{\prime }} \\
\partial _{A_2^{\prime }} \\
... \\
\partial _{A_z^{\prime }}
\end{array}
\right) =\left(
\begin{array}{c}
\frac \partial {\partial x^{I^{\prime }}} \\
\frac \partial {\partial y_{(1)}^{A_1^{\prime }}} \\
\frac \partial {\partial y_{(2)}^{A_2^{\prime }}} \\
... \\
\frac \partial {\partial y_{(z)}^{A_z^{\prime }}}
\end{array}
\right)
$$
and
$$
\widehat{{\bf Y}}{\bf =}\widehat{Y}_{<\alpha >}^{<\alpha ^{\prime }>}=\left(
\begin{array}{ccccc}
K_I^{I^{\prime }} & Y_{(1,0)I}^{A_1^{\prime }} & Y_{(2,0)I}^{A_2^{\prime }}
& ... & Y_{(z,0)I}^{A_z^{\prime }} \\
0 & K_{A_1}^{A_1^{\prime }} & Y_{(2,1)A_1}^{A_2^{\prime }} & ... &
Y_{(z,1)A_1}^{A_z^{\prime }} \\
0 & 0 & K_{A_2}^{A_2^{\prime }} & ... & Y_{(z,2)A_2}^{A_z^{\prime }} \\
... & ... & ... & ... & ... \\
0 & 0 & 0 & ... & K_{A_z}^{A_z^{\prime }}
\end{array}
\right) .
$$

We note that we obtain a supersimmetric generalization of the
Miron--Atanasiu \cite{mirata} osculator bundle $\left( Osc^z\tilde M,\pi
,\tilde M\right) $ if the fiber space is taken to be a direct sum of $z$
vector s-spaces of the same dimension $\dim {\cal F}=\dim \widetilde{M},$
i.e. ${\cal F}^{<d>}={\cal F}\oplus {\cal F}\oplus ...\oplus {\cal F}.$ In
this case the $K$ and $Y$ matrices from (3) and (6) satisfy identities:%
$$
K_{A_1}^{A_1^{\prime }}=K_{A_2}^{A_2^{\prime }}=...=K_{A_z}^{A_z^{\prime }},
$$
$$
Y_{(1,0)A}^{A^{\prime }}=Y_{(2,1)A}^{A^{\prime
}}=...=Y_{(z,z-1)A}^{A^{\prime }},
$$
$$
.............................
$$
$$
Y_{(p,0)A}^{A^{\prime }}=Y_{(p+1,1)A}^{A^{\prime
}}=...=Y_{(z,z-1)A}^{A^{\prime }},\quad (p=2,...,z-1).
$$
For $z=1$ the $Osc^1\widetilde{M}$ is the ts-bundle $T\widetilde{M}.$

Introducing projection $\pi _0^z\doteq \pi ^{<z>}:\widetilde{{\cal E}}%
^{<z>}\rightarrow \widetilde{M}$ we can also consider projections $\pi
_{p_2}^{p_1}:\widetilde{{\cal E}}^{<p_1>}\rightarrow \widetilde{{\cal E}}%
^{<p_2>}~\quad (p_2<p_1)$ defined as
$$
\pi _{s_2}^{s_1}(x,y^{(1)},...,y^{(p_1)})=(x,y^{(1)},...,y^{(p_2)}).
$$
The s-differentials $d\pi _{p_2}^{p_1}:T(\widetilde{{\cal E}}%
^{<p_1>})\rightarrow T(\widetilde{{\cal E}}^{<p_2>})$ of maps $\pi
_{p_2}^{p_1}$ in turn define vertical dvs-subbundles $V_{h+1}=Kerd\pi
_h^{p_1}~(h=0,1,..,p_1-1)$ of the tangent dvs-bundle $T(\widetilde{{\cal E}}%
^{<z>})~$( the dvs-space $V_1=V$ is the vertical dvs-subbundle on $%
\widetilde{{\cal E}}^{<z>}.$ The local fibres of dvs-subbundles $V_h$
determines this regular s-distribution $V_{h+1}:u\in \widetilde{{\cal E}}%
^{<z>}\rightarrow V_{h+1}(u)\subset T(\widetilde{{\cal E}}^{<z>})$ for which
one holds inclusions $V_z\subset V_{z-1}\subset ...\subset V_1.\,$ The
enumerated properties of vertical dvs--subbundles are explicitly illustrated
by transformation laws (6) for distinguished local bases.

\section{Nonlinear Connections in Vector S--Bundles}

The purpose of this section is to present an introduction into geometry of
the nonlinear connection structures in vector s-bundles. The concept of
nonlinear connection (N-connection) was introduced in the fra\-me\-work of
Finsler geometry \cite{car,car35,kaw} (the global definition of
N-connec\-ti\-on is given in \cite{barth}). It should be noted that the
N-connection (splitting) field could play an important role in modeling
various variants of dynamical reduction from higher dimensional to lower
dimensional s--spaces with (or not) different types of local anisotropy.
Monographs \cite{ma87,ma94} containes details on geometrical properties of
N-connection structures in v--bundles and different generalizations of
Finsler geometry and some proposals (see Chapter XII in \cite{ma87}, written
by S. Ikeda) on physical interpretation of N--connection in the framework of
''unified'' field theory with interactions nonlocalized by y--dependencies
are discussed. We emphasize that N-connection is a different geometrical
object from that introduced by using nonlinear realizations of gauge groups
and supergroups (see, for instance, the collection of works on supergravity
\cite{sal} and approaches to gauge gravity \cite{ts,pon}). To make the
presentation to aid rapid assimilation we shall have realized our geometric
constructions firstly for vs-bundles then (in the next section) we shall
extend them for higher order extensions, i.e. for general dvs--bundles.

Let consider the definitions of N--connection structure \cite{vlasg} in a
vs-bundle $\tilde {{\cal E}}=(\tilde E,{\pi }_E,\tilde M)$ whose typical
fibre is ${\hat {{\cal F}}}$ and ${{\pi }^T}:T\tilde {{\cal E}}{\to }T\tilde
M$ is the superdifferential of the map ${{\pi }_E}$ (${{\pi }^T}$ is a
fibre-preserving morphism of the ts-bundle $(T\tilde {{\cal E}},{{\tau }_E}%
,\tilde M)$ to $\tilde E$ and of ts-bundle $(T\tilde M,{\tau },\tilde M)$ to
$\tilde M$). The kernel of this vs-bundle morphism being a subbundle of $%
(T\tilde E,{{\tau }_E},\tilde E)$ is called the vertical subbundle over $%
\tilde {{\cal E}}$ and denoted by $V\tilde {{\cal E}}=(V\tilde E,{{\tau }_V}%
,\tilde E)$. Its total space is $V\tilde {{\cal E}}={{\bigcup }_{u\in \tilde
{{\cal E}}}}{\quad }{V_u},{\quad }$ where ${V_u}={ker}{{\pi }^T},{\quad }{u{%
\in }\tilde {{\cal E}}}.$ A vector
$$
Y={Y^\alpha }{\frac{{\partial }}{{{\partial }{u^\alpha }}}}={Y^I}{\frac{{%
\partial }}{{{\partial }{x^I}}}}+{Y^A}{\frac{{\partial }}{{{\partial }{y^A}}}%
}={Y^i}{\frac{{\partial }}{{{\partial }{x^i}}}}+{Y^{\hat i}}{\frac{{\partial
}}{{{\partial }{{\theta }^{\hat i}}}}}+{Y^a}{\frac{{\partial }}{{{\partial }{%
y^a}}}}+{Y^{\hat a}}{\frac \partial {\partial {{\zeta }^{\hat a}}}}
$$
tangent to $\tilde {{\cal E}}$ in the point $u\in \tilde {{\cal E}}$ is
locally represented as
$$
(u,Y)=({u^\alpha },{Y^\alpha })=({x^I},{y^A},{Y^I},{Y^A})=({{\hat x}^i},{{%
\theta }^{\hat i}},{{\hat y}^a},{{\zeta }^{\hat a}},{{\hat Y}^i},{Y^{\hat i}}%
,{{\hat Y}^a},{Y^{\hat a}}).
$$

A nonlinear connection, N-connection, in vs-bundle $\tilde {{\cal E}}$ is a
splitting on the left of the exact sequence

$$
0{\longmapsto }{V\tilde {{\cal E}}}\stackrel{i}{\longmapsto }{T\tilde {{\cal %
E}}}{\longmapsto }{{T\tilde {{\cal E}}{/}V\tilde {{\cal E}}}}{\longmapsto }0,%
\eqno(7)
$$
i.e. a morphism of vs-bundles $N:T\tilde {{\cal E}}\in {V\tilde {{\cal E}}}$
such that $N{\circ }i$ is the identity on $V\tilde {{\cal E}}$.

The ker\-nel of the mor\-phism $N$ is called the hor\-i\-zon\-tal
sub\-bun\-dle and denoted by
$$
(H\tilde E,{{\tau }_E},\tilde E).
$$
From the exact sequence (7) one follows that N-connection structure can be
equivalently defined as a distribution $\{{{{\tilde E}_u}\to {{H_u}\tilde E}}%
,{{T_u}\tilde E}={{H_u}\tilde E}{\oplus }{{V_u}\tilde E}\}$ on $\tilde E$
defining a global decomposition, as a Whitney sum,
$$
T{\tilde {{\cal E}}}=H{\tilde {{\cal E}}} \oplus V{\tilde {{\cal E}}}.%
\eqno(8)
$$

To a given N-connection we can associate a covariant s-derivation on ${%
\tilde M}:$

$$
{\bigtriangledown }_X{Y}=X^I{\{{\frac{{\partial Y^A}}{{\partial x^I}}}+{N_I^A%
}(x,Y)\}}s_A,\eqno(9)
$$
where $s_A$ are local independent sections of $\tilde {{\cal E}},{\quad }Y={%
Y^A}s_A$ and $X={X^I}s_I$.

S-differentiable func\-tions $N^{A}_{I}$ from (3) writ\-ten as func\-tions
on $x^I$ and $y^{A},\\ N^{A}_{I}(x,y),$ are called the coefficients of the
N-connection and satisfy these transformation laws under coordinate
transforms (1) in ${\cal E}$:
$$
N^{A^{\prime}}_{I^{\prime}}{\frac{{\partial x^{I^{\prime}}}}{{\partial x^{I}}%
}}=M^{A^{\prime}}_{A} N^{A}_{I}- {\frac{\partial {M^{A^{\prime}}_{A}{(x)}}}{%
\partial x^I}} {y^A}.%
$$

If coefficients of a given N-connection are s-differentiable with respect to
coordinates $y^A$ we can introduce (additionally to covariant nonlinear
s-derivation (9)) a linear covariant s-derivation $\hat D$ (which is a
generalization for vs-bundles of the Berwald connection \cite{berw}) given
as follows:
$$
{{\hat D}_{({\frac{{\partial }}{{\partial x^I}}})}}({\frac{{\partial }}{{%
\partial y^A}}})={{{\hat N}^B}_{AI}}({\frac{{\partial }}{{\partial y^B}}}),{%
\quad }{{\hat D}_{({\frac{{\partial }}{{\partial y^A}}})}}({\frac{{\partial }%
}{{\partial y^B}}})=0,
$$
where

$$
{{\hat N}^A}_{BI}(x,y)={\frac{{{\partial }{{N^A}_I}{(x,y)}}}{{\partial y^B}}}%
\eqno(10)
$$
and
$$
{{{\hat N}^A}_{BC}}{(x,y)}=0.
$$
For a vector field on ${\tilde {{\cal E}}}{\quad }Z={Z^I}{\frac \partial {{%
\partial x^I}}}+{Y^A}{\frac \partial {{\partial y^A}}}$ and $B={B^A}{(y)}{%
\frac \partial {{\partial y^A}}}$ being a section in the vertical s-bundle $%
(V\tilde E,{{\tau }_V},\tilde E)$ the linear connection (10) defines
s-derivation (compare with (9)):
$$
{{\hat D}_Z}B=[{Z^I}({\frac{\partial B^A}{\partial x^I}}+{\hat N}%
_{BI}^AB^B)+Y^B{\frac{\partial B^A}{\partial y^B}}]{\frac \partial {\partial
y^A}}.
$$

Another important characteristic of a N-connection is its curvature:
$$
\Omega ={\frac 12}{\Omega }_{IJ}^A{dx^I}\land {dx^J}\otimes {\frac \partial
{\partial y^A}}
$$
with local coefficients
$$
{\Omega }_{IJ}^A={\frac{\partial N_I^A}{\partial x^J}}-{(-)}^{|IJ|}{\frac{%
\partial N_J^A}{\partial x^I}}+N_I^B{\hat N}_{BJ}^A-{(-)}^{|IJ|}N_J^B{\hat N}%
_{BI}^A,\eqno(11)
$$
where for simplicity we have written ${(-)}^{{\mid K\mid }{\mid J\mid }}={(-)%
}^{\mid {KJ}\mid }.$

We note that lin\-ear con\-nec\-tions are par\-tic\-u\-lar cases of
N-connections locally pa\-ra\-met\-rized as $N_I^A{(x,y)}=N_{BI}^A{(x)}%
x^Iy^B,$ where functions $N_{BI}^A{(x)}$, defined on $\tilde M,$ are called
the Christoffel coefficients.

\section{N-Connections in DVS-Bundles}

In order to define a N-connection in a dvs--bundle $\widetilde{{\cal E}}%
^{<z>}$ we consider a s--sub\-bund\-le $N\left( \widetilde{{\cal E}}%
^{<z>}\right) $ of the ts-bundle $T\left( \widetilde{{\cal E}}^{<z>}\right) $%
for which one holds (see \cite{sau} and \cite{mirata} respectively for jet
and osculator bundles) the Whitney sum (compare with (8))%
$$
T\left( \widetilde{{\cal E}}^{<z>}\right) =N\left( \widetilde{{\cal E}}%
^{<z>}\right) \oplus V\left( \widetilde{{\cal E}}^{<z>}\right) .
$$
$N\left( \widetilde{{\cal E}}^{<z>}\right) $ can be also interpreted as a
regular s--distribution (horizontal distribution being supplementary to the
vertical s-distribution $V\left( \widetilde{{\cal E}}^{<z>}\right) )$
determined by maps $N:u\in \widetilde{{\cal E}}^{<z>}\rightarrow N(u)\subset
T_u\left( \widetilde{{\cal E}}^{<z>}\right) .$

The condition of existence of a N-connection in a dvs-bundle $\widetilde{%
{\cal E}}^{<z>}$ can be proved as in \cite{ma87,ma94,mirata}: It is required
that $\widetilde{{\cal E}}^{<z>}$ is a paracompact s--differentiable (in our
case) manifold.

Locally a N--connection in $\widetilde{{\cal E}}^{<z>}$ is given by its
coefficients
$$
N_{(01)I}^{A_1}(u),(N_{(02)I}^{A_2}(u),
N_{(12)A_1}^{A_2}(u)),...(N_{(0p)I}^{A_p}(u),N_{(1p)A_1}^{A_p}(u),...
N_{(p-1p)A_{p-1}}^{A_p}(u)),...,
$$
$$
(N_{(0z)I}^{A_z}(u),N_{(1z)A_1}^{A_z}(u),...,N_{(pz)A_p}^{A_z}(u),...,N_{(z-1z)A_{z-1}}^{A_z}(u)),
$$
where, for instance, $%
(N_{(0p)I}^{A_p}(u),N_{(1p)A_1}^{A_p}(u),...,N_{(p-1p)A_{p-1}}^{A_p}(u))$
are components of N--connection in vs-bundle $\pi ^{<p>}:\tilde
E^{<p>}\rightarrow \tilde E^{<p-1>}.$ Here we note that if a
N-con\-nect\-i\-on structure is defined we must correlate to it the local
partial derivatives on $\widetilde{{\cal E}}^{<z>}$ by considering instead
of local coordinate bases (4) and (5) the so--called locally bases
(la-bases)
$$
\delta _{<\alpha >}=(\delta _I,\delta _{<A>})=(\delta _I,\delta
_{(A_1)},\delta _{(A_2)},...,\delta _{(A_s)})=
$$
$$
\frac \delta {\partial u^{<\alpha >}}=(\frac \delta {\partial x^I},\frac
\delta {\partial y_{(1)}^{A_1}},\frac \delta {\partial
y_{(2)}^{A_2}},...,\frac \delta {\partial y_{(z)}^{A_z}})\eqno(12)
$$
(the dual la--bases are denoted as%
$$
\delta ^{<\alpha >}=(\delta ^I,\delta ^{<A>})=(\delta ^I,\delta
^{(A_1)},\delta ^{(A_2)},...,\delta ^{(A_z)})=
$$
$$
\delta u^{<\alpha >}=(\delta x^I,\delta y^{(A_1)},\delta
y^{(A_2)},...,\delta y^{(A_z)})\quad )\eqno(13)
$$
with components parametrized as
$$
\delta _I=\partial _I-N_I^{A_1}\partial _{A_1}-N_I^{A_2}\partial
_{A_2}-...-N_I^{A_{z-1}}\partial _{A_{z-1}}-N_I^{A_z}\partial _{A_z},%
\eqno(14)
$$
$$
\delta _{A_1}=\partial _{A_1}-N_{A_1}^{A_2}\partial
_{A_2}-N_{A_1}^{A_3}\partial _{A_3}-...-N_{A_1}^{A_{z-1}}\partial
_{A_{z-1}}-N_{A_1}^{A_z}\partial _{A_z},
$$
$$
\delta _{A_2}=\partial _{A_2}-N_{A_2}^{A_3}\partial
_{A_3}-N_{A_2}^{A_4}\partial _{A_4}-...-N_{A_2}^{A_{z-1}}\partial
_{A_{z-1}}-N_{A_2}^{A_z}\partial _{A_z},
$$
$$
..............................................................
$$
$$
\delta _{A_{z-1}}=\partial _{A_{z-1}}-N_{A_{z-1}}^{A_z}\partial _{A_z},
$$
$$
\delta _{A_z}=\partial _{A_z},
$$
or, in matrix form, as%
$$
{\bf \delta }_{\bullet }{\bf =}\widehat{{\bf N}}(u)\times {\bf \partial }%
_{\bullet },
$$
where%
$$
{\bf \delta }_{\bullet }{\bf =}\delta _{<\alpha >}=\left(
\begin{array}{c}
\delta _I \\
\delta _{A_1} \\
\delta _{A_2} \\
... \\
\delta _{A_z}
\end{array}
\right) =\left(
\begin{array}{c}
\frac \delta {\partial x^I} \\
\frac \delta {\partial y_{(1)}^{A_1}} \\
\frac \delta {\partial y_{(2)}^{A_2}} \\
... \\
\frac \delta {\partial y_{(z)}^{A_z}}
\end{array}
\right) ,{\bf \partial }_{\bullet }{\bf =}\partial _{<\alpha >}=\left(
\begin{array}{c}
\partial _I \\
\partial _{A_1} \\
\partial _{A_2} \\
... \\
\partial _{A_z}
\end{array}
\right) =\left(
\begin{array}{c}
\frac \partial {\partial x^I} \\
\frac \partial {\partial y_{(1)}^{A_1}} \\
\frac \partial {\partial y_{(2)}^{A_2}} \\
... \\
\frac \partial {\partial y_{(z)}^{A_z}}
\end{array}
\right)
$$
and%
$$
\widehat{{\bf N}}{\bf =}\left(
\begin{array}{ccccc}
1 & -N_I^{A_1} & -N_I^{A_2} & ... & -N_I^{A_z} \\
0 & 1 & -N_{A_1}^{A_2} & ... & -N_{A_1}^{A_z} \\
0 & 0 & 1 & ... & -N_{A_2}^{A_z} \\
... & ... & ... & ... & ... \\
0 & 0 & 0 & ... & 1
\end{array}
\right) \eqno(15)
$$
In generalized index form we write the matrix (6) as $\widehat{N}_{<\beta
>}^{<\alpha >},$ where, for instance, $\widehat{N}_J^I=\delta _J^I,\widehat{N%
}_{B_1}^{A_1}=\delta _{B_1}^{A_1},...,\widehat{N}_I^{A_1}=-N_I^{A_1},...,%
\widehat{N}_{A_1}^{A_z}=-N_{A_1}^{A_z},\widehat{N}%
_{A_2}^{A_z}=-N_{A_2}^{A_z},...~.$

So in every point $u\in \widetilde{{\cal E}}^{<z>}$ we have this invariant
decomposition:%
$$
T_u\left( \widetilde{{\cal E}}^{<d>}\right) =N_0(u)\oplus N_1(u)\oplus
...\oplus N_{z-1}(u)\oplus V_z(u),
$$
where $\delta _I\in N_0,\delta _{A_1}\in N_1,...,\delta _{A_{z-1}}\in
N_{z-1},\partial _{A_z}\in V_z.$

We note that for the osculator s-bundle $\left( Osc^z\tilde M,\pi ,\tilde
M\right) $ there is an additional (we consider the N-adapted variant)
s--tangent structure
$$
J:\Xi \left( Osc^z\tilde M\right) \rightarrow \Xi \left( Osc^z\tilde
M\right)
$$
defined as
$$
\frac \delta {\partial y_{(1)}^I}=J\left( \frac \delta {\partial x^I}\right)
,...,\frac \delta {\partial y_{(z-1)}^I}=J\left( \frac \delta {\partial
y_{(z-2)}^I}\right) ,\frac \partial {\partial y_{(z)}^I}=J\left( \frac
\delta {\partial y_{(z-1)}^I}\right) \eqno(16)
$$
(in this case $I$- and $A$-indices take the same values and we can not
distinguish them), by considering vertical $J$-distributions
$$
N_0=N,N_1=J\left( N_0\right) ,...,N_{z-1}=J\left( N_{z-2}\right) .
$$
In consequence, for the la-adapted bases on $\left( Osc^z\tilde M,\pi
,\tilde M\right) $ there is written this N--connection matrix:%
$$
{\bf N=}N_{<I>}^{<J>}=\left(
\begin{array}{ccccc}
1 & -N_{(1)I}^J & -N_{(2)I}^J & ... & -N_{(z)I}^J \\
0 & 1 & -N_{(1)I}^J & ... & -N_{(z-1)I}^J \\
0 & 0 & 1 & ... & -N_{(z-2)I}^J \\
... & ... & ... & ... & ... \\
0 & 0 & 0 & ... & 1
\end{array}
\right) .\eqno(17)
$$

There is a unique distinguished local decomposition of every s--vector $X\in
\chi \left( \widetilde{{\cal E}}^{<z>}\right) $ on la-base (12):%
$$
X=X^{(H)}+X^{(V_1)}+...+X^{(V_z)},\eqno(18)
$$
by using the horizontal, $h,$ and verticals, $v_1,v_2,...,v_z,$ projections:%
$$
X^{(H)}=hX=X^I\delta _I,~X^{(V_1)}=v_1X=X^{(A_1)}\delta
_{A_1},...,X^{(V_z)}=v_zX=X^{(A_z)}\delta _{A_z}.
$$

With respect to coordinate transforms (3) the la-bases (12) and ds--vector
components (18) are correspondingly transformed as
$$
\frac \delta {\partial x^I}=\frac{{\partial }x^{I^{\prime }}}{{\partial }x^I}%
\frac \delta {\partial x^{I^{\prime }}},~\frac \delta {\partial
y_{(p)}^{A_p}}=K_{A_p}^{A_p^{\prime }}\frac \delta {\partial
y_{(p)}^{A_p^{\prime }}},\eqno(19)
$$
and%
$$
X^{I^{\prime }}=\frac{{\partial }x^{I^{\prime }}}{{\partial }x^I}%
X^I,~X^{(A_p^{\prime })}=K_{A_p}^{A_p^{\prime }}X^{(A_p^{\prime })},\forall
p=1,2,...,z.
$$

Under changing of coordinates (3) the local coefficients of a nonlinear
connection transform as follows:%
$$
Y_{<\alpha >}^{<\alpha ^{\prime }>}\widehat{N}_{<\alpha ^{\prime }>}^{<\beta
^{\prime }>}=\widehat{N}_{<\alpha >}^{<\beta >}(K_{<\beta >}^{<\beta
^{\prime }>}+Y_{<\beta >}^{<\beta ^{\prime }>})
$$
(we can obtain these relations by putting (19) and (6) into (14) where $%
\widehat{N}_{<\alpha ^{\prime }>}^{<\beta ^{\prime }>}$ satisfy $\delta
_{<\alpha ^{\prime }>}=\widehat{N}_{<\alpha ^{\prime }>}^{<\beta ^{\prime
}>}\partial _{<\beta ^{\prime }>}).$

For dual la-bases (13) we have these N--connection ''prolongations of
differentials'':%
$$
\delta x^I=dx^I,
$$
$$
\delta y^{A_1}=dy_{(1)}^{A_1}+M_{(1)I}^{A_1}dx^I,\eqno(20)
$$
$$
\delta
y^{A_2}=dy_{(2)}^{A_2}+M_{(2)A_1}^{A_2}dy_{(1)}^{A_1}+M_{(2)I}^{A_2}dx^I,
$$
$$
................................
$$
$$
\delta
y^{A_s}=dy_{(s)}^{A_s}+M_{(s)A_1}^{As}dy_{(1)}^{A_1}+M_{(s)A_2}^{As}dy_{(2)}^{A_2}+...+M_{(z)I}^{A_z}dx^I,
$$
where $M_{(\bullet )\bullet }^{\bullet }$ are the dual coefficients of the
N-connection which can be expressed explicitly by recurrent formulas through
the components of N--connection $N_{<A>}^{<I>}.$ To do this we shall rewrite
formulas (20) in matrix form:%
$$
{\bf \delta }^{\bullet }={\bf d}^{\bullet }\times {\bf M}(u),
$$
where
$$
{\bf \delta }^{\bullet }=\left(
\begin{array}{ccccc}
\delta x^I & \delta y^{A_1} & \delta y^{A_2} & ... & \delta y^{A_s}
\end{array}
\right) ,~{\bf d}^{\bullet }=\left(
\begin{array}{ccccc}
dx^I & dy_{(1)}^{A_1} & \delta y_{(2)}^{A_2} & ... & \delta y_{(s)}^{A_s}
\end{array}
\right)
$$
and%
$$
{\bf M=}\left(
\begin{array}{ccccc}
1 & M_{(1)I}^{A_1} & M_{(2)I}^{A_2} & ... & M_{(z)I}^{A_z} \\
0 & 1 & M_{(2)A_1}^{A_2} & ... & M_{(z)A_1}^{A_z} \\
0 & 0 & 1 & ... & M_{(z)A_2}^{A_z} \\
... & ... & ... & ... & ... \\
0 & 0 & 0 & ... & 1
\end{array}
\right) ,
$$
and then, taking into consideration that bases ${\bf \partial }_{\bullet }(%
{\bf \delta }_{\bullet })$ and ${\bf d}^{\bullet }({\bf \delta }^{\bullet })$
are mutually dual, to compute the components of matrix ${\bf M}$ being
s--inverse to matrix ${\bf N}$ (see (15)). We omit these simple but tedious
calculus for general dvs-bundles and, for simplicity, we present the basic
formulas for osculator s--bundle $\left( Osc^z\tilde M,\pi ,\tilde M\right) $
when $J$--distribution properties (16) and (17) alleviates the problem. For
common type of indices on $\tilde M$ and higher order extensions on $%
Osc^z\tilde M$ the dual la-base is expressed as
$$
\delta x^I=dx^I,
$$
$$
\delta y_{(1)}^I=dy_{(1)}^I+M_{(1)J}^Idx^J,\eqno(21)
$$
$$
\delta y_{(2)}^I=dy_{(2)}^I+M_{(1)J}^Idy_{(1)}^J+M_{(2)J}^Idx^J,
$$
$$
................................
$$
$$
\delta
y_{(z)}^I=dy_{(z)}^I+M_{(1)J}^Idy_{(s-1)}^J+M_{(2)J}^Idy_{(z-2)}^J+...+M_{(z)J}^Idx^J,
$$
with $M$--coefficients computed by recurrent formulas:%
$$
M_{(1)J}^I=N_{(1)J}^I,
$$
$$
M_{(2)J}^I=N_{(2)J}^I+N_{(1)K}^IM_{(1)J}^K,
$$
$$
..............
$$
$$
M_{(s)J}^I=N_{(s)J}^I+N_{(s-1)K}^IM_{(1)J}^K+...+N_{(2)K}^IM_{(z-2)J}^K+N_{(1)K}^IM_{(z-1)J}^K.
$$
One holds these transformation law for dual coefficients (21) with respect
to coordinate transforms (3):
$$
M_{(1)J}^KY_{(0,0)K}^{I^{\prime }}=M_{(1)K^{\prime }}^{I^{\prime
}}Y_{(0,0)J}^{K^{\prime }}+Y_{(1,0)J}^{I^{\prime }},
$$
$$
M_{(2)J}^KY_{(0,0)K}^{I^{\prime }}=M_{(2)K^{\prime }}^{I^{\prime
}}Y_{(0,0)J}^{K^{\prime }}+M_{(1)K^{\prime }}^{I^{\prime
}}Y_{(1,0)J}^{K^{\prime }}+Y_{(2,0)J}^{I^{\prime }},
$$
$$
................................
$$
$$
M_{(z)J}^KY_{(0,0)K}^{I^{\prime }}=M_{(z)K^{\prime }}^{I^{\prime
}}Y_{(0,0)J}^{K^{\prime }}+M_{(z-1)K^{\prime }}^{I^{\prime
}}Y_{(1,0)J}^{K^{\prime }}+...+M_{(1)K^{\prime }}^{I^{\prime
}}Y_{(z-1,0)J}^{K^{\prime }}+Y_{(z,0)J}^{I^{\prime }}.
$$
(the proof is a straightforward regroupation of terms after we have put (3)
into (21)).

Finally, we note that curvatures of a N-connection in a dvs-bundle $%
\widetilde{{\cal E}}^{<z>}$ can be introduced in a manner similar to that
for usual vs-bundles (see (11) by a consequent step by step inclusion of
higher dimension anisotropies :
$$
\Omega _{(p)}={\frac 12}{\Omega }_{(p)\alpha _{p-1}\beta _{p-1}}^{A_p}{%
\delta u}^{\alpha _{p-1}}\land {\delta u}^{\beta _{p-1}}\otimes \frac \delta
{\partial y_{(p)}^{A_p}},~p=1,2,...,z,
$$
with local coefficients
$$
{\Omega }_{(p)\beta _{p-1}\gamma _{p-1}}^{A_p}=\frac{\delta N_{\beta
_{p-1}}^{A_p}}{\partial u_{(p-1)}^{\gamma _{p-1}}}-{(-)}^{|\beta
_{p-1}\gamma _{p-1}|}\frac{\delta N_{\gamma _{p-1}}^{A_p}}{\partial
u_{(p-1)}^{\beta _{p-1}}}+
$$
$$
N_{\beta _{p-1}}^{D_p}{\hat N}_{D_p\gamma _{p-1}}^{A_p}-{(-)}^{|\beta
_{p-1}\gamma _{p-1}|}N_{\gamma _{p-1}}^{D_p}{\hat N}_{D_p\beta _{p-1}}^{A_p},%
$$
where ${\hat N}_{D_p\gamma _{p-1}}^{A_p}=\frac{\delta N_{\gamma _{p-1}}^{A_p}%
}{\partial y_{(p)}^{D_p}}$ (we consider $y^{A_0}\simeq x^I).$

\section{Discussion}

We have explicitly constructed a new class of superspaces with higher order
anisotropy. The status of the results in this work and the relevant open
questions are discussed as follows.

From the generally mathematical point of view it is possible a definition of
a supersymmetric differential geometric structure imbedding both type of
supersymmetric extensions of Finsler and Lagrange geometry as well various
Kaluza--Klein superspaces. The first type of superspaces, considered as
locally anisotropic, are characterized by nontrivial nonlinear connection
structures and corresponding distinguishing of geometric objects and basic
structure equations. The second type as a rule is associated to trivial
nonlinear connections and higher order dimensions. A substantial interest
for further considerations presents the investigations of physical
consequences of models of field interactions on higher and/or lower
dimensional superspaces provided with N--connection structure.

It worth noticing that higher order derivative theories are one of currently
central division in modern theoretical and mathematical physics. It is
necessary a rigorous formulation of the geometric background for developing
of higher order analytic mechanics and corresponding extensions to classical
quantum field theories. Our results do not only contain a supersymmetric
extension of higher order fiber bundle geometry, but also propose a general
approach to the ''physics'' with locally anisotropic interactions. The
elaborated in this paper formalism of distinguished vector superbundles
highlights a scheme by which supergravitational and superstring theories
with higher order anisotropy can be constructed. This is a matter of our
further investigations \cite{vlasg,vlags}.

\vskip20pt {\bf Acknowledgments}

The author is much obliged to Profs R. Miron and U. Bruzzo for reprints of
their papers.

\end{document}